\documentclass{article}
\usepackage{graphicx}

\begin{document}

\title{On the Effectiveness of Defensive Distillation}
\author{Nicolas Papernot and Patrick McDaniel\\ The Pennsylvania State University\\\{ngp5056,mcdaniel\}@cse.psu.edu}

\maketitle

\begin{abstract}
\noindent We report experimental results indicating that defensive distillation successfully mitigates adversarial samples crafted using the fast gradient sign method~\cite{goodfellow2014explaining}, in addition to those crafted using the Jacobian-based iterative attack~\cite{papernot2015limitations} on which the defense mechanism was originally evaluated. 
\end{abstract}

\section{Introduction}
Despite their successful application to several machine learning tasks, deep neural networks have been shown to be vulnerable~\cite{szegedy2013intriguing}, upon completion of their training, to evasion attacks---as is the case with other machine learning techniques~\cite{biggio2013evasion,papernot2016transferability}. In the case of deep neural networks, such attacks can be performed by crafting \emph{adversarial examples}~\cite{szegedy2013intriguing}: subtly (and often humanly indistinguishably) modified malicious inputs crafted to compromise the integrity of the deep neural network outputs. Following~\cite{biggio2013evasion,szegedy2013intriguing}, several approaches have been introduced to craft adversarial examples~\cite{goodfellow2014explaining,papernot2015limitations}. We refer  readers interested in the details of these crafting algorithms to the following book chapter~\cite{warde2016adversarial}, which includes a detailed presentations of these different approaches. In this technical report, we discuss a defense mechanism proposed in~\cite{papernot2015distillation}, named \emph{defensive distillation}. Specifically, we present experimental results complementary to the evaluation of the attack's effectiveness that was included in the original paper~\cite{papernot2015distillation}.

\section{Defensive Distillation}

\noindent Traditionally, a deep neural network $f$ used for classification is trained on a dataset $(X,Y)=\{(x,y)\}$, which is a collection of pairs $(x,y)$ of inputs $x$ and class labels $y$. A class label $y$ simply indicates the index of the  class to which sample $x$ belongs and can be encoded as an indicator vector. The neural network is trained to minimize its  prediction error $\|f(x)-y\|$ by back-propagating~\cite{lecun1998gradient} the gradients of a cost function---quantifying the error made on each sample $x$ in the training dataset $X$---with respect to the neural network's parameters.\\

\noindent Introduced in~\cite{papernot2015distillation} as a defense mechanism to increase the robustness of deep neural networks, \emph{defensive distillation} alters the training of a deep neural network used for classification essentially by re-configuring its last layer. This last layer---a \emph{softmax}---computes probabilities $f_i(x)$ for each class $i$ of the problem by applying the following transformation to the logits, which can be considered as class scores $z(x)$ produced by the neural network: 
\begin{equation}
f_i(x) =\frac{e^{z_i(x)/T}}{\sum_{l=0}^{n-1} e^{z_l(x)/T}}
\end{equation}
The inputs of the softmax layer are thus divided by a factor $T$ named \emph{temperature}. This temperature parameter is traditionally set to $T=1$ in the literature. Higher temperatures $T$ yield uniformly distributed probabilities: each class is assigned a probability close to $1/n$ where $n$ is the number of classes in the classification problem. Instead, lower temperatures yield discrete probability vectors with a large probability (close to $1$) assigned to the most likely class and small probabilities (close to $0$) assigned to the other classes.\\

\noindent  Defensive distillation~\cite{papernot2015distillation} alters the training of a classifier in the following way:
\begin{enumerate}
	\item A first instance of the neural network $f$ is trained as described above using the training data $(X,Y)$ where the labels $Y$ indicate the correct class of samples $X$. However, the temperature of the neural network softmax is raised to high values---larger than $1$.
	\item This first neural network instance is used to infer predictions on the training data. This produces a new training dataset $(X,f(X))$ where the class labels indicating the correct class of samples $X$ are replaced with probability vectors quantifying the likeliness of $X$ being in each class. 
	\item A second---\emph{distilled}---instance of the neural network $f$ is trained using this newly labeled dataset $(X,f(X))$. The distilled network is trained at the same high temperature than the first instance. However, when training completes and the network is deployed to make predictions, its temperature is set back to $1$ so as to increase the confidence of its predictions. 
\end{enumerate} 

\section{Robustness of Distilled Neural Networks to the Fast Gradient Sign Method}

In~\cite{papernot2015distillation}, defensive distillation is evaluated on the attack introduced in~\cite{papernot2015limitations}, which estimates the sensitivity of the targeted deep neural network by computing its Jacobian and iteratively altered components with large sensitivity. The evaluation of distillation showed that the attack, originally successful at rates of $97\%$ on MNIST~\cite{lecun1998mnist} was mitigated, reducing the success rate to $0.45\%$. We now present an evaluation of defensive distillation as a defense mechanism to mitigate adversarial examples crafted using the fast gradient sign method, introduced in~\cite{goodfellow2014explaining}.\\

\begin{figure}
    \centering
    \includegraphics[width=\textwidth]{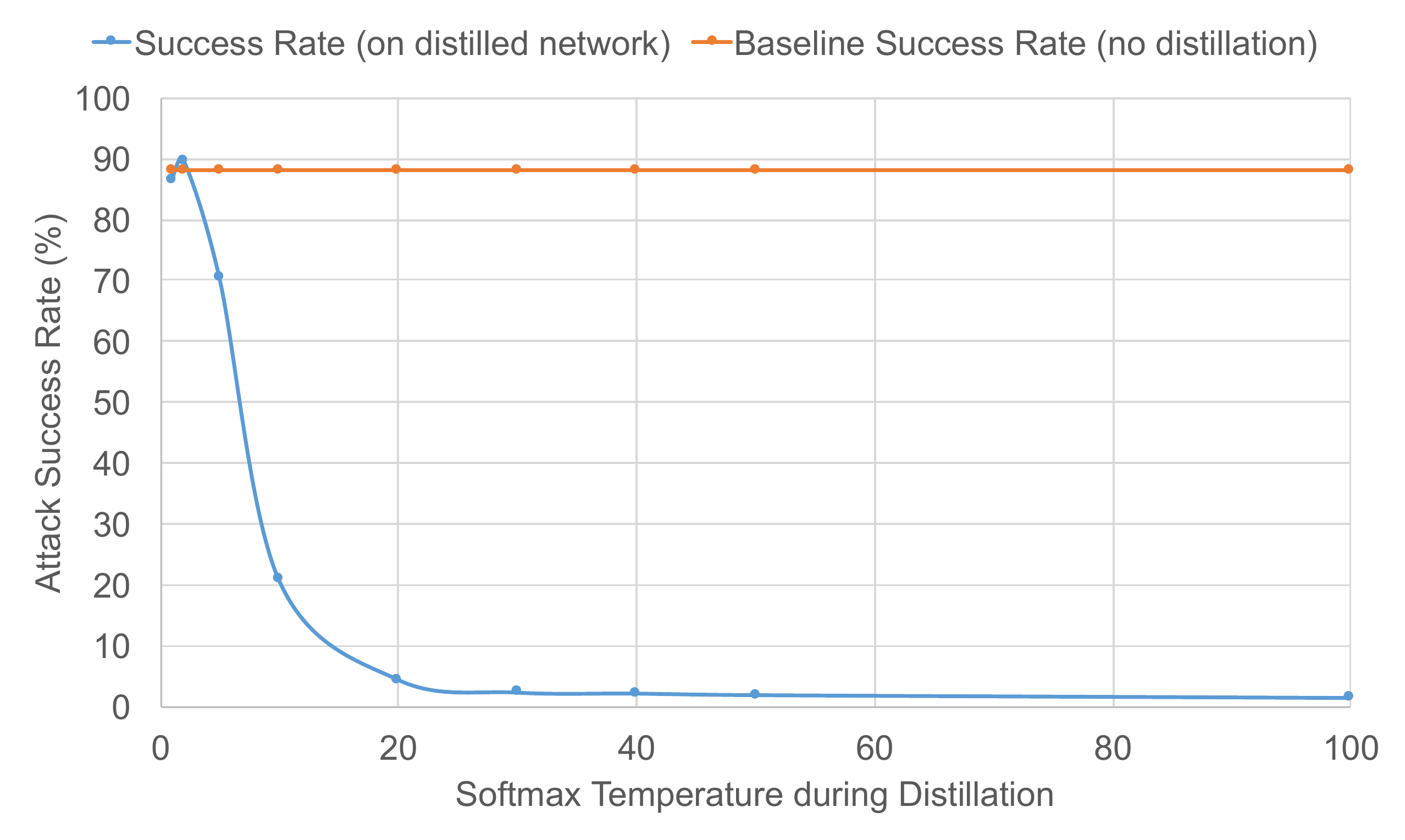}
    \caption{Success of the Fast Gradient Sign Method~\cite{goodfellow2014explaining} when defensive distillation is applied to the MNIST model, as described in~\cite{papernot2015distillation}.}
    \label{figure}
\end{figure}

\noindent  We reproduce the experimental setup described in~\cite{papernot2015distillation}. It considers a $9$ layer deep neural network classifying handwritten digits of the MNIST dataset. We train a baseline network without distillation. It achieves $99.51\%$ accuracy on the test set, which is comparable to state-of-the-art performance on this task. This network can be attacked using the fast gradient sign method~\cite{goodfellow2014explaining} with a success rate of $88.03\%$ when the input variation parameter is set to $\varepsilon=0.3$. We then train a collection of distilled networks for several temperatures $T$ ranging from $1$ to $100$, and measure the success of the attack for each of these networks by crafting an adversarial example for each of the $10,000$ test set samples. The results are reported in Figure~\ref{figure}. As temperature increases, the attack is mitigated with a success rate smaller than $1.5\%$ at a temperature of $T=100$.

\section{Conclusion}
We empirically demonstrated that defensive distillation mitigates adversarial samples crafted using the fast gradient sign method introduced in~\cite{goodfellow2014explaining}, in addition to those crafted using the Jacobian-based iterative approach introduced in~\cite{papernot2015limitations}.

\newpage

\bibliographystyle{abbrv}
\bibliography{dl-biblio}

\end{document}